\documentclass[]{IET}

\usepackage{cite}
\usepackage{algorithm}
\usepackage{algorithmic}
\renewcommand{\algorithmicrequire}{\textbf{Input:}}
\renewcommand{\algorithmicensure}{\textbf{Output:}}

\usepackage{graphicx}
\usepackage{upgreek}
\usepackage{graphics}

\usepackage{float}
\usepackage{subfigure}
\usepackage{array}
\usepackage{multirow} 
\usepackage{amsmath}
\usepackage{xcolor}
\usepackage{booktabs}
\usepackage{url}
\usepackage{lettrine}

\begin{document}

\title{Predictive and Recommendatory Spectrum Decision for Cognitive Radio}

\author{Xinran Chen}
\author{Zhe Chen\thanks{We conceived the main idea of this paper in 2012. The Chinese manuscript of this paper was finished in May, 2015.}}

\author{Sai Xie}

\author{Yongshuai Shao}

\affil{School of Computer Science and Engineering, Northeastern University, China}
\affil{Email: chenzhe@mail.neu.edu.cn}

\abstract{Cognitive radio technology enables improving the utilization efficiency of the precious and scarce radio spectrum. How to maximize the overall spectrum efficiency while minimizing the conflicts with primary users is vital to cognitive radio. The key is to make the right decisions of accessing the spectrum. Spectrum prediction can be employed to predict the future states of a spectrum band using previous states of the spectrum band, whereas spectrum recommendation recommends secondary users a subset of available spectrum bands based on secondary user's previous experiences of accessing the available spectrum bands. In this paper, a framework for spectrum decision based on spectrum prediction and spectrum recommendation is proposed. As a benchmark, a method based on extreme learning machine (ELM) for single-user spectrum prediction and a method based on Q-learning for multiple-user spectrum prediction are proposed. At the stage of spectrum decision, two methods based on Q-learning and Markov decision process (MDP), respectively, are also proposed to enhance the overall performance of spectrum decision. Experimental results show that the performance of the spectrum decision framework is much better.}

\maketitle

\section{Introduction}

In recent years, due to the rapid development of wireless communication, the number of wireless communication equipment and the demand of wireless communication increase gradually, the spectrum resources become more precious and scare.
Cognitive radio (CR) technology has been put forward to make efficient use of the scarce radio frequency spectrum to increase the overall spectrum efficiency. With intelligence and cognitive abilities, cognitive radio is able to find out spectrum holes and make use of the spectrum holes through spectrum sensing and dynamic spectrum access technologies. Spectrum decision is the ability of CR to select the best available spectrum band for secondary users (SUs), without causing harmful interference to primary users (PUs). To improve the overall utilization and throughput of spectrum bands, best effort should be exerted to reduce the probability of collision between SUs and PUs.

In this paper, we focus on spectrum decision of cognitive radio.
We propose a spectrum decision framework based on spectrum prediction and spectrum recommendation. Then we investigate the methods for spectrum prediction and spectrum recommendation, respectively. For single-user spectrum prediction, a method based on extreme learning machine (ELM) is proposed.
For multiple secondary users, a Q-learning based collaborative spectrum prediction method is proposed.
For spectrum recommendation, a method based on the cooperative filtering is applied.


This paper is organized as follows: the related work is discussed in Section 2. In Section 3, prediction and recommendation based spectrum decision model are presented with detailed description. Q-learning and ELM based spectrum prediction algorithms are proposed in Section 4. A spectrum prediction method based on collaborative filtering recommendation is proposed in section 5. In Section 6, the design of spectrum decision method based on prediction and recommendation is presented. Then, spectrum decision modeling and algorithm based on prediction and recommendation is proposed in section 7. In Section 8, the performance analysis and evaluation of our methods are presented. Finally, Section 9 concludes this paper.

\section{Related Work}

In 2005, Simon Haykin put forward the early concept of cognitive radio spectrum prediction in~\cite{haykin2005cognitive}. Since 2007, more and more methods for spectrum prediction have emerged, such as the binary time series method, auto-regressive model, Markov model, Nueral network, and Bayesian networks model. A collaborative spectrum sensing and prediction method which can decrease secondary user's interference to primary user is proposed in \cite{9}. Paper \cite{10} proposes an auto-regressive based spectrum prediction model. And in \cite{11} a particle filter based auto-regressive channel model is proposed, which performs better than that of \cite{10}. A hidden Markov based adaptive channel state prediction model is proposed in \cite{13}.
Paper \cite{Chen2010tvt} proposes an advanced Markov chain based single-user channel-state prediction algorithm as well as a collaborative prediction algorithm for multiple SUs.
Experimental results show that the performance of advanced Markov chain based spectrum prediction is better than that of the nearest neighbor prediction method, and the collaborative prediction algorithm also outperforms the M-out-of-N algorithm. In \cite{16,17,18}, a backward propagation (BP) neural network based spectrum prediction algorithm is proposed. In \cite{19,20} a differential evolution and Levenberg-Marquardt based spectrum prediction algorithm is proposed, which improves the accuracy of the BP based spectrum prediction method. A support vector machine (SVM) based spectrum prediction algorithm is proposed in \cite{21}, and its performance is better than that of BP neural network. In \cite{22}, the author proposes a feedback neural network based spectrum prediction algorithm. In this method, power sample value instead of channel state is used as the input of spectrum prediction. Experimental results show that the performance is better than traditional methods. In \cite{23} a time varying non-stationary hidden Markov model based spectrum prediction is proposed with enhanced performance.

In recent years, recommendation technology has been introduced to the field of cognitive radio. In 2010, recommendation system was firstly introduced to cognitive radio by Li and validated in \cite{25}. Paper \cite{26} proposes to apply collaborative filtering to cognitive radio, for secondary users to choose suitable channels. A modified collaborative filtering algorithm which considers the location of each secondary user is proposed in \cite{27}. In \cite{28}, the author presents a dynamic feature model for secondary users according to the theory of interacting particle systems. In \cite{29}, the problem of channel recommendation is described as an average reward based Markov decision process, and a model reference adaptive search method is proposed. In a word, most of the existing researches on spectrum recommendation just focus on collaborative based spectrum recommendation methods. And few of them consider the scenario of multiple channels with many PUs and SUs.

Spectrum decision is fundamental to CR. It is the ability of a cognitive radio to select the best available spectrum band for SUs without causing harmful interferences to PUs \cite{32}. To improve the overall utilization and throughput of communication channels, cognitive radio should do its best to reduce the probability of collision between SUs and PUs.

Much research for spectrum decision and spectrum allocation has been done in recent years. In paper \cite{30}, a spectrum decision framework is proposed to determine a set of spectrum bands by considering application requirements as well as the dynamic nature of spectrum bands. Paper \cite{31} proposes a selective opportunistic spectrum access (SOSA) scheme. With the aid of statistical data and traffic prediction techniques, the SOSA scheme can estimate the probability of a channel appearing idle based on the statistics and choose the best spectrum-sensing order to maximize spectrum efficiency and maintain an SU’s connection. Paper \cite{32} provides a survey on spectrum decision in CR networks and addresses issues of spectrum characterization, spectrum selection, and CR reconfiguration. Paper \cite{33} combines the sensing and prediction to enhance the spectrum utilization and reduce interferences to PUs. And an effective solution is proposed using the dual optimal theory. Paper \cite{34} proposes a dynamic spectrum access scheme where secondary users cooperatively recommend ``good" channels to each other and access accordingly. In paper \cite{35}, a graph-theoretical model is developed to characterize different traffic demands between CR users by using interfere graph, on the basis of traditional labeling system. They propose a traffic-demand algorithm based on the graph theoretical model, which can support different traffic demands of CR users that change with time. Paper \cite{36} makes an attempt to study such wireless networks with opportunistic spectrum availability and access. Paper \cite{39} proposes a multichannel selection algorithm that uses spectrum hole prediction to limit the interference to primary networks and to exploit channel characteristics in order to enhance channel utilization. Paper \cite{40} evaluates the use of supervised machine learning for channel selection in wireless sensor networks. In paper \cite{41} , a new efficient Taguchi algorithm based on orthogonal arrays (OA) is proposed to deal with the 0-1 discrete spectrum allocation optimization problem.

However, all the existing works failed to consider the problem of spectrum prediction, spectrum recommendation, and spectrum decision as a whole. And they failed to connect spectrum prediction with spectrum recommendation. In this paper, we regard spectrum recommendation as a complement to spectrum prediction and integrate both of them to the subsequent spectrum decision.

\section{Problem Formulation}

 In this section, a basic channel state and slot division model is firstly introduced. In order to enhance the accuracy of spectrum decision, both spectrum prediction and spectrum recommendation are utilized to support spectrum decision. The a model of spectrum decision based on both spectrum prediction and spectrum recommendation is proposed.

\subsection{Channel State and Slot Division Model}


Suppose the licensed spectrum of primary users can be divided into $M$ channels, and PUs in every two channels are independent from each other. The channel state of primary user can be divided into two types, i.e., busy and idle. Let ``OFF" denote the idle channel and ``ON" denote the busy channel. $S_{i}(t)$ denotes the channel state $i$ at specific time slot $t$. The channel state can be defined as follows:
\begin{equation}
S_{i}\!=\!\begin{cases}
0, OFF(idle)\\
1, ON(busy)\\
\end{cases}
\end{equation}

The most widely used channel state model is ON-OFF model~\cite{42} which continuously alternates between ON and OFF. There are two main streams of ways for building channel model. One is queuing model~\cite{43}, the other is Markov state model~\cite{44}. In \cite{45} Geirhofer and Tong built an actual signal based mathematical model by extracting signal features in 802.11b WLAN of 2.4GHz ISM frequency band. Because the distribution of the channel state of primary users conforms to queue system model, the queuing system model is selected to simulate the usage of the channel of primary user in this paper.

The queuing model can be denoted as $X/Y/Z$ and the definition of $X,Y,Z$ are expressed as follows:\\
\begin{enumerate}
\item[X:] The distribution of the time of customer arrival.
\item[Y:] The distribution of the duration of the customer accepting the service.
\item[Z:] The number of available servers.
\end{enumerate}
In this paper, the customer means PU and the server means frequency band or channel.
The arrival time of primary users conforms to Poisson distribution and the duration of primary user occupying a channel conforms to geometric distribution. The Poisson and geometric distributions are expressed as follows:
\begin{equation}
    P_{n}(t)=\frac{e^{-\lambda t}(\lambda t)^{n}}{n!}\qquad P(X=k)=(1-p)^{k-1}p
\end{equation}
\begin{figure}[!t]
	\centering
	\includegraphics[width=5.5in]{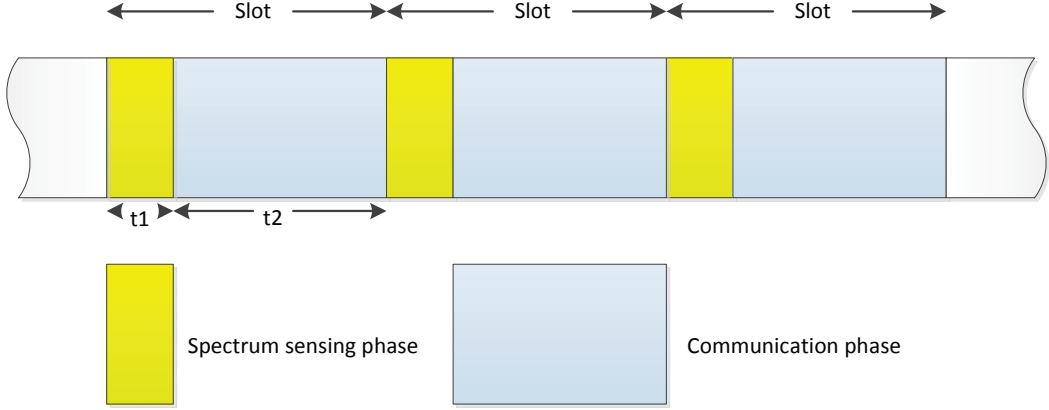}
	\caption{Time slots in cognitive radio}
	\label{fig:fig1}
\end{figure}
In cognitive radio, in order to avoid disturbing primary users, secondary users should continuously sensing the channel of primary users. Time slot is usually used in cognitive radio. And a channel can be divided into many slots~\cite{Chen2010tvt} as shown in Fig.~\ref{fig:fig1}.

Each slot consists of two phases. The first phase is for spectrum sensing. In this phase, SUs detect the state of a channel state (``busy" or ``idle"). The second phase is for communication. If the result of spectrum sensing is ``idle", SUs can occupy the channel to communication in this phase. Otherwise, SUs continue to sense the channel. From Fig.~1 we can learn that the length of each slot $t_{slot}$ can be denotes as $t_{slot}=t_{1}+t_{2}$.

\subsection{Prediction and Recommendation based Spectrum Decision Model}

 \begin{figure}[!htbp]
	\centering
	\includegraphics[width=6in]{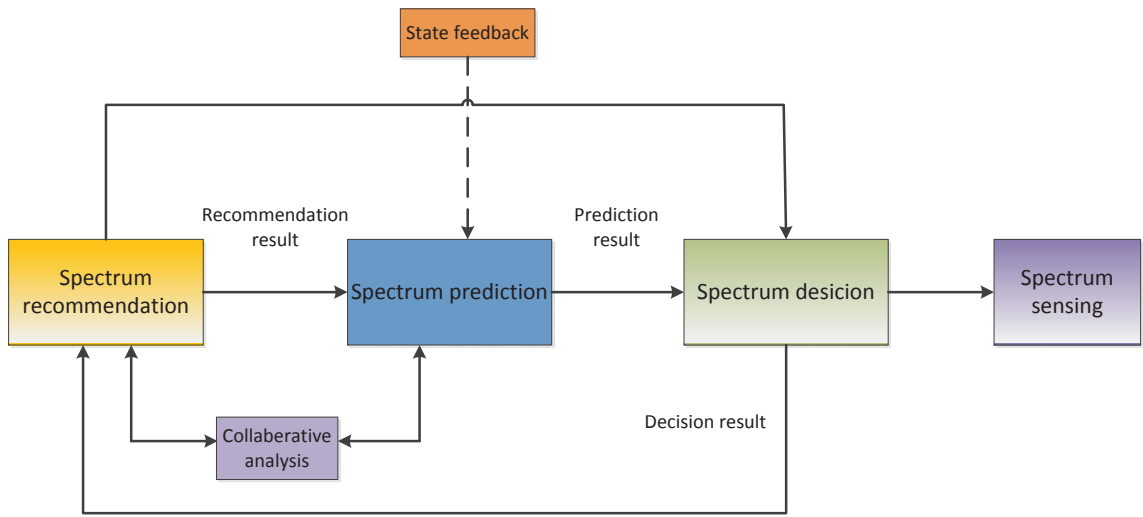}
	\caption{Spectrum decision model based on spectrum prediction and spectrum recommendation}
	\label{fig:fig2}
\end{figure}
 \begin{figure}[!htbp]
	\centering
	\includegraphics[width=4.5in]{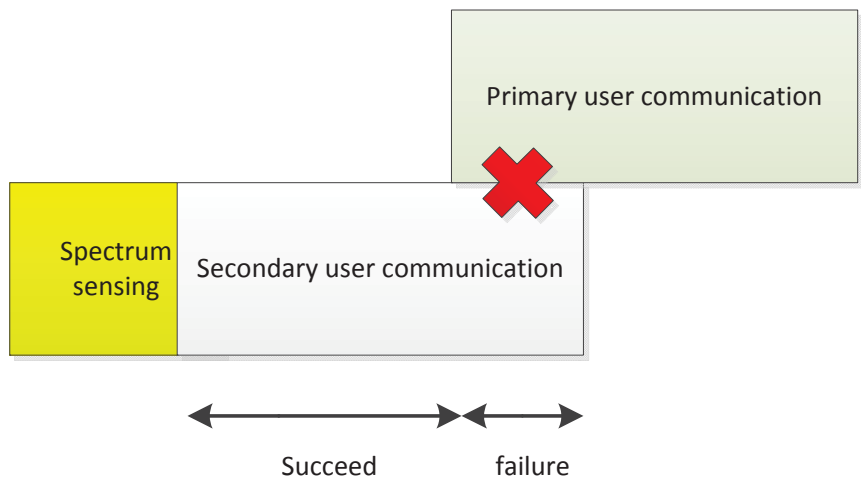}
	\caption{Collision between primary user and second user}
	\label{fig:fig3}
\end{figure}

Fig.~\ref{fig:fig2} shows prediction and recommendation based spectrum decision model. The design idea and the relationship among each module are introduced as follows.

The key research for cognitive radio system is to maximize the utilization of spectrum resource and avoid collisions between users. The purpose for spectrum prediction and recommendation is to select optimal channel and decrease the collision rate among users. Although spectrum prediction and recommendation are all the basis for channel selection, however, they are different from each other. Spectrum prediction predicts future channel state using the history information of the channel acquired by spectrum sensing. And spectrum recommendation recommends a better channel according to the experiences of secondary users who have accessed this channel before.

A previous research applied spectrum prediction to spectrum decision and allocation. In \cite{29} the spectrum prediction and sensing is applied to spectrum decision and the total throughput is increased. To our best knowledge, there is no research on using both spectrum prediction and recommendation for spectrum decision.
Thus, a unified model for spectrum decision incorporating both spectrum prediction and spectrum recommendation is proposed in this paper.

In Fig.~\ref{fig:fig2}, the proposed model consists of spectrum prediction, spectrum recommendation, spectrum correlation, and spectrum decision. The function of each module and their relationship are described as follows.

Spectrum prediction module: Use the history information of the channel state of primary users to predict the future state of the channel (busy or idle).

Spectrum recommendation module: Use the information of experiences of secondary users in accessing the channels to recommend a better channel for future access.

Spectrum decision module: According to the outputs of the spectrum prediction module and the spectrum recommendation module, a final decision for access which channel is made in this module.

\section{Spectrum Prediction}

In this section, we focus on spectrum prediction. First of all, the extreme learning machine based spectrum prediction is proposed. Then a cooperative spectrum prediction method based on Q-learning is also proposed.

\subsection{Extreme Learning Machine for Spectrum Prediction}

Extreme Learning Machine(ELM) was proposed by Guangbin Huang in 2004. It is a kind of single hidden layer feedforward neural network(SLFN). It randomly chooses the input weights and analytically determines the output weights of SLFNs. The learning speed can be thousands of times faster than traditional feedforward network learning algorithms like back-propagation algorithm while obtaining better generalization performance. Suppose there are S samples $(x_{i},t_{i})\in R^{d_{i}} \times R^{d_{2}}$, $d_{1}$ denotes the size of input, $d_{2}$ denotes the size of output. The number of hidden layer node is L, where $L \leq S$, $g(\cdot)$ is activation function. $(w_{i},b_{i})$ denotes the connection weights and threshold, $\beta_{i}$ are the output weights. Single layer feedforward neural network can be expressed as follows.
\begin{equation}\sum_{i=1}^{L}\beta_{i}G(w_{i},b_{i},x_{i})=t_{j}, j=1,2,\cdots,S\end{equation}
And it can be translated into matrix style:
\begin{equation}H\beta=T\end{equation}
where the matrix $H$ is:
\begin{equation}
H=
\left[\begin{array}{ccc}
G(w_{1},b_{1},x_{1}) & \cdots & G(w_{L},b_{L},x_{1}) \\
\vdots & \cdots & \vdots \\
G(w_{1},b_{1},x_{s}) & \cdots & G(w_{L},b_{L},x_{s})
\end{array}
\right]
\end{equation}
where $T$ and $\beta$ is:
\begin{equation}
T=
 \begin{bmatrix}
   t_{1}^{T} \\
   \vdots  \\
   t_{S}^{T}
  \end{bmatrix},
\beta=
  \begin{bmatrix}
   \beta_{1}^{T} \\
   \vdots  \\
   \beta_{S}^{T}
  \end{bmatrix}
\end{equation}
$H$ denotes the output matrix of hidden neural nodes. The column $i$ denotes the output of $x_{1},x_{2}\cdots,x_{S}$ as the input of $i^{th}$ hidden neural node. Actually, the parameter for input neural nodes is initialized randomly, and the process of network training is to get the output weights $\beta$ through least square solution.
\begin{equation}H\beta=H^{+}T=(H^{T}H)^{-1}H^{T}T\end{equation}
ELM algorithm consists of the following three steps:\\
For a giving training set $D=\{(x_{i},t_{i}),i=l,\cdots,S\}$, activation function $g(x)$, and the number of hidden layer $L$.\\
\begin{enumerate}
\item[*] Generate the output weights and bias randomly $(w_{i},b_{i}), i=1,\cdots,L$.
\item[*] Compute the output matrix $H$ for hidden layer nodes.
\item[*] Compute weight $\beta$: $\beta=H^{+}T$.
\end{enumerate}
\begin{figure}[!t]
	\centering
	\includegraphics[width=3in]{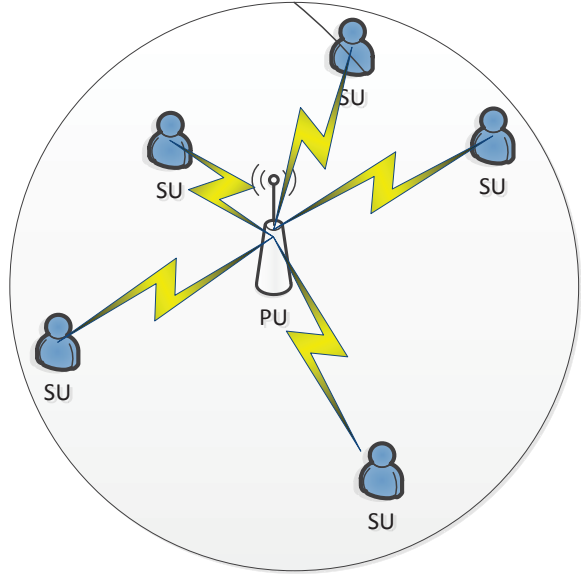}
	\caption{Illustration of the scenario of spectrum prediction}
	\label{fig:system scen}
\end{figure}

Suppose there are $N$ secondary users, i.e., $SU_{1},SU_{2},\cdots,SU_{N}$, and one primary user $PU$, as shown in Fig.~\ref{fig:system scen}. The primary user occupies one licensed channel. The secondary users do not collaborate. With previous states of the channel acquired by spectrum sensing, secondary users can predict the next channel state.

Spectrum prediction uses history data of channel state, i.e., $\{s_{1},s_{2},\cdots,s_{t-1},s_{t}\}$, to predict the channel state of next time slot $s_{t+1}$. The state of the time slot close to $s_{t+1}$ is more valuable for prediction. Thus, we select $n$ slots $\{s_{t-n},s_{t-n+1},\cdots,s_{t-1},s_{t}\}$ as the input to the ELM based spectrum prediction. And $s_{t+1}$ is the predicted channel state.

The process of ELM neural network based spectrum prediction is as follows.

1. Build the model of ELM neural network. The number of neural nodes in input layer, hidden layer, and output layer are $n, L, m$, respectively. The $n$ channel states of time slots  $\{s_{t-n+1},s_{t-n+1},\cdots,s_{t}\}$ are input to the input layer. And the number of hidden layer is determined by serval experiments. ELM is different from BP neural network, as it needs to set the number of hidden layer nodes. The weights and threshold are initialized randomly.

2. Train the ELM neural network spectrum prediction model. Use history channel state data as the training data. ELM is different from other neural network, in the process of training, since it does not need to set learning speed and max training time. It only needs to select a suitable activation function. We choose sine as the activation function.

3. ELM neural network spectrum prediction. With the model trained by previous step, the test data $\{s_{t-n+1},s_{t-n+1},\cdots,s_{t-1},s_{t}\}$ are used as the input of ELM. And the output of ELM is regarded as predicted state. Then, the difference between the actual states and predicted states of the channel is analyzed.

\subsection{Q-learning Based Spectrum Prediction Algorithm}

Q-learning is a form of model-free reinforcement learning. It provides agents with the capability of learning to act optimally in Markovian domains by experiencing the consequences of actions, without requiring them to build maps of the domains.
Q-learning $Q(s,\alpha)$ can make optimal decisions with the state of $s$ and a simplified decision process. The basic formula of Q-learning is shown as follows.
\begin{equation}Q^{*}(s,\alpha)=R(s,\alpha)+\gamma\sum_{s\in S}T(s,\alpha,s^{'})max_{\alpha^{'}}Q^{*}(s^{'},\alpha^{'})\end{equation}
Where $Q^{*}(s,\alpha)$ is the expected discounted reward for executing action $\alpha$ at
state $s$, and $\gamma$ is the discount factor. The goal of Q-learning is to estimate the $Q$ values for an optimal policy. 

In the process of Q-learning, the table of $Q$ value is updated continuously. The agent can decide its optimal action at each time $t$, according to condition state $s_{t}$, and observe the rewards value $r$ in new conditional state $s_{t+1}$. The update formula is shown as Eq. 9.
\begin{equation}Q(s_{t},\alpha_{t})=Q(s_{t},\alpha_{t})+\alpha[r_{t+1}+\gamma maxQ(s_{t+1},\alpha_{t+1})-Q(s_{t},\alpha_{t})]\end{equation}

The steps of each iteration for Q-learning are as follows.
\begin{enumerate}
\item[1)] Initialize $Q(s,\alpha)$ and its parameters such as $\alpha$,$\gamma$, and $t=0$.
\item[2)] Observe current condition state $s_{t}$.
\item[3)] According to current $Q(s,\alpha)$, select $\alpha_{t}$ which can maximize $Q$ value.
\item[4)] Observe the next condition state $s_{t+1}$ and the rewards value $r_{t+1}$ after
finishing the action $\alpha_{t}$.
\item[5)] Use Eq. 9 to update the value of $Q(s,a)$.
\item[6)] If current state is the target state, stop iteration. Otherwise, go to step 3, and let $t=t+1$.
\end{enumerate}

Cooperative spectrum prediction is a process of predicting the state of primary user's channel and it is based on the prediction results of each participant secondary user. The cooperative spectrum prediction can improve the accuracy of spectrum prediction. In \cite{62} a Q-learning based cooperative spectrum sensing approach is proposed. Q-learning is a reinforcement learning method which is good at solving dynamic decision problem. One of the traditional methods named M-out-of-N is naturally suitable for hard combination of multiple decisions. But it lacks the ability of autonomous learning. Autonomous learning ability is one of the advantages of Q-learning. A method for cooperative spectrum prediction based on Q-learning is proposed in this section.


Suppose there are $N$ secondary users predicting the channel state of primary user independently. The local prediction result of each secondary user is denoted as $s_{i,t}$, which is the prediction result of $SU_{i}$ at slot $t$, where $i \in \{1,2,\cdots N\}$, $s_{i,t}\in \{0,1\}$. The prediction results $s_{i,t}$ of each $SU_{i}$ can be regarded as one condition state of Q-learning. Different combination of local prediction result can form different condition state. $\widetilde{s_{t}}$ denotes the Q-learning state at time $t$. The $N$ prediction results of $SU_{i}$ are $N$ binary numbers. Let ``0" denote the ``idle" channel state and ``1" denote the ``busy" channel state. The $N$ one-bit binary numbers can form an integer $\widetilde{s_{t}}$ whose value ranges from $0$ to $2^{N}-1$.
 \begin{equation}\widetilde{s_{t}}=\sum_{i=1}^{N}s_{i,t}\times2^{i-1}\end{equation}

The output action $\alpha_{t}$ of Q-learning is the result of the cooperative spectrum prediction, which is the predicted channel state of primary user, $\alpha_{t}\in \{0,1\}$.
$\alpha_{t}=0$ means the channel is predicted to be ``idle", whereas $\alpha_{t}=1$ means the channel is predicted to be ``busy".

Assigning values to rewards $r$ for Q-learning can be tricky. If the result of cooperative spectrum prediction $\alpha_{t}$ equals to the actual state $s_{t}^{\alpha}$, then $r$ is a reward value, otherwise $r$ is a penalty value. In the proposed method, $r$ is assigned as follows.
\begin{equation}
r\!=\!\begin{cases}
R_{p} &\alpha_{t}=s_{t}^{\alpha}\\
R_{n} &\alpha_{t} \neq s_{t}^{\alpha}\\
\end{cases}
\end{equation}
where $\alpha_{t}$ is the result of cooperative spectrum prediction at slot $t$. $s_{t}^{\alpha}$ is the actual channel state of primary user. $R_{r}$ and $R_{p}$ are constants.

\renewcommand{\algorithmicrequire}{\textbf{Input:}}
\renewcommand{\algorithmicensure}{\textbf{Output:}}
\begin{algorithm} 
\caption{Q-learning based spectrum prediction algorithm} 
\label{alg1} 
\begin{algorithmic}[1] 
\STATE Build Q-table and initialize parameters, such as learning rate $\alpha$, discount factor $\gamma$, and decision time $t=0$.
\STATE Each participant secondary user predicts the channel state of primary user independently, and sends the prediction result $s_{i,t}$ to a central node.
\STATE According to the prediction result of each secondary user, the condition state $\widetilde{s_{t}}$ can be figured out.
$\widetilde{s_{t}}=\sum_{i=1}^{N}s_{i,t}\times2^{i-1}$
\STATE Choose an action $\alpha_{t}$ which can maximize the Q value at current state. $\alpha_{t}=max_{\alpha}Q(s,\alpha)$
\STATE Compare to the actual channel state $c_{t}$ and evaluate the prediction result. If $\alpha_{t}=c_{t}$, the reward value is set to $R_{p}$. Otherwise it is set to $R_{n}$.
\STATE Update $Q(s,\alpha)$
$Q(s_{t},\alpha_{t})=Q(s_{t},\alpha_{t})+\alpha[r_{r+1}+\gamma maxQ(s_{t+1},\alpha_{t+1})-Q(s_{t},\alpha_{t})]$
\STATE If the final goal is achieved, then stop. Otherwise go to step 2 and $t=t+1$.
\end{algorithmic}
\end{algorithm}

\section{Collaborative Filtering Recommendation based Spectrum Prediction Method}

Spectrum recommendation is a recent technology which is proposed in recent years. Collaborative filtering is one of the recommendation methods. In this section, a recommendation system based on collaborative filtering for cognitive radio is introduced.

\subsection{Collaborative Filtering Recommendation Algorithm}

\begin{figure}[!t]
	\centering
	\includegraphics[width=6in]{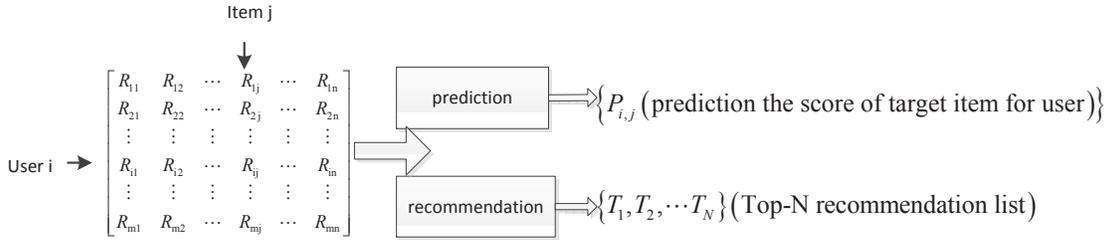}
	\caption{Collaborative filtering recommendation system}
	\label{fig:Method figure of collaborative filtering system}
\end{figure}

Recommendation system is used to suggest new items or to predict the utility of a certain item for a particular user based on the user's previous likings and
the opinions of other like-minded users. The basic idea of collaborative filtering algorithm is to provide a recommendation list for target users based on the opinions of other like-minded users.
The figure of collaborative filtering recommendation system is shown in Fig.~\ref{fig:Method figure of collaborative filtering system}.

There are two types of collaborative filtering based recommendation algorithms, i.e., item-based and user-based collaborative filter algorithms. Both of them are based on user-item score matrix to build recommendation system model. We use item-based collaborative filter algorithms.

Item-based algorithms use the set of items related to the target user to compute how similar they are to the target item $j$ and select $top-N$ most similar items $\{j_{1},j_{2},\cdots,j_{N}\}$. Then the similarity between $top-N$ items and the target item $j$ are computed $\{P_{i1},P_{i2},\cdots,P_{iN}\}$. After working out the most similar items, we can give a prediction result by taking a weighted average of the target $top-N$ similar items.

\subsection{Generating Recommendation List}

For a target user $u$, the neighbor set $N(u)={u_{1},u_{2},\cdots,u_{k}}$ can be obtained by the methods mentioned above. A recommendation result can be generated by this neighbor set. First of all, the score of a specific item for target user can be work out. Secondly, target $top-N$ recommendation list is generated. A prediction value $P_{ui}$ can be obtained by the average score of neighbor set for a specific item.
\begin{equation}P_{ui}=\frac{1}{K}\sum_{v\in N(u)}R_{vi}\end{equation}

Eq. 12 shows that all users in neighbor set are equal treated. But this is usually not practical. Because different similarity causes different degree of effect on prediction. Thus, Eq. 13 shows an improve method. It computes the prediction on an item $i$ for a user $u$ by computing the sum of the ratings given by the user on the items similar to $i$. Each rating is weighted by the corresponding similarity between $u$ and items.
\begin{equation} P_{ui}=\frac{\sum_{v\in N(u)}sim(u,v)\cdot R_{vi}}{\sum_{v\in N(u)}\mid sim(u,v)\mid}\end{equation}

Eq. 14 shows a method which considers the fact that different user has scoring deviation. It improves the prediction accuracy.
\begin{equation}P_{ui}=\frac{\sum_{v\in N(u)}sim(u,v)\cdot (R_{vi}-\bar{R_{v}})}{\sum_{v\in N(u)}\mid sim(u,v)\mid}+\bar{R_{u}}\end{equation}
where $\bar{R_{u}},\bar{R_{v}}$ denotes the average score of user $u$ and $v$. Let $I_{u}=\{i\in I\mid R_{ui}\neq 0\}$ denote all of the items that user $u$ has already given the score. Then $R_{u}$ can be expressed as follows.
\begin{equation} \bar{R_{u}}=(1/\mid I_{u}\mid)\sum_{i\in I_{u}}R_{ui} \end{equation}

\subsection{Collaborative Filtering Based Spectrum Recommendation Algorithm}

Collaborative filtering algorithm is a commonly used recommendation algorithm. 
Collaborative filtering algorithms mainly focus on the relationship between users and items. In cognitive radio, secondary users can be regarded as users and channels can be regarded as items. However, if collaborative filtering algorithm is used for cognitive radio spectrum recommendation, there are still some problems to be solved.

In cognitive radio, time dimension can not be ignored. Because the preference of one specific user can not be changed frequently, so in a collaborative filtering algorithm the user's score for an item can not be changed in a limited time. However, in cognitive radio, channel state changes frequently, the sensing result of secondary user for channel state has strong relationship with time. Thus, when collaborative filtering is used in spectrum recommendation, the recommendation list must be updated in real-time.

The score matrix for $N$ secondary users and $M$ primary users can be described as $R_{MN}$. $R_{ij}$ is the score of user $i$ for channel $j$.
\begin{figure}[!t]
	\centering
	\includegraphics[width=4in]{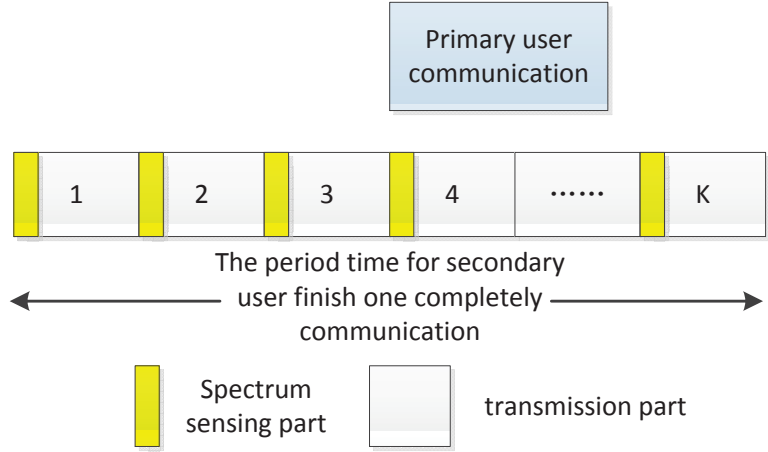}
    \caption{Illustration of scoring rules}
	\label{fig:Illustration of scoring rules}
\end{figure}

The scoring value is based on the number of transmission slots for primary users before primary users access the channel. Thus, if the score $R_{ij}$ is high, it denotes that the primary user transmits a large quantity of data. Fig.~\ref{fig:Illustration of scoring rules} shows that the primary user starts communication after the secondary user transmits for $3$ time slots. Then, the score value is three $R_{ij}=3$. The initial score matrix can be obtained by secondary users accessing primary user's channel randomly over a period of time.

In spectrum prediction, each secondary user has the identical feature. 
Thus, the similarity between two secondary users is $sim(u,v)=1$. All of the secondary users are from the neighbor set. And based on this assumption, the score of other secondary users for accessing the channel is the same as that of current secondary user accessing the channel.

The state of channel varies with time. The channel state can be regards to be constant in very short time. Thus, the average score $finalscore_{j}$ of nearest $L$ secondary users $SU_{i}$ for a random channel $channel_{j}$ can be regarded as the score of target user for this channel.
\begin{equation}finalscore_{j}=\frac{1}{Total_{t}}\sum_{L}^{Total}R_{ij}\end{equation}
where $Total_{t}$ is the total number of nearest $L$ time slots starting at time $t$ when secondary users access channel $channel_{j}$.
If the value of $finalscore$ is higher than a preset threshold $Th$, then $channel_{j}$ is the recommended channel. If the number of secondary users waiting for accessing the channel is less than the number of users in the recommendation list, the secondary users access the channel in the same order of sorting $finalscore$ from high to low.

\section{The Design of Spectrum Decision Method Based on Prediction and Recommendation}
\subsection{Design Idea and Purpose}

Existing spectrum allocation methods are mainly based on spectrum sensing or spectrum prediction. This is because that spectrum prediction can predict future spectrum holes through the history usage information of PU's channel. Optimizing the prediction algorithm can reduce SUs' interference on PUs. Spectrum prediction focuses on PUs' action and behavior. On the contrary, spectrum recommendation cares more about SUs' user experience. Therefore, we propose to combine spectrum prediction and spectrum recommendation to further reduce the collision between users. Fig.~2 shows the framework of spectrum decision based on prediction and recommendation.

Different from traditional spectrum allocation, the dynamic allocation is the significant characteristic of cognitive radio. Reinforcement learning has the self-learning ability and can make decisions dynamically based on self-learning status. By modeling based on reinforcement learning, intelligent and dynamic spectrum decision can be realized.

\subsection{Spectrum Decision System Model}

In practical situations, communications between users are conducted in pairs in our system model. However, due to the limitations of communications distance, SUs may not communicate with others when their geographical distance is beyond communication range. Besides, as the distance increases, the similarity between SUs will decrease. That will influence the results of spectrum recommendations, thus spectrum decision. Therefore, our spectrum decision system model considers two scenarios. One is that all SUs are within the communication range and can communicate with each other. The other is that SUs are scattered, so SUs can only communicate with others that are within its communication range.

\textbf{Scenario One}: SUs' communication distance and locations are ignore.

Assume that PUs and SUs are randomly distributed within the scope of a cognitive radio system. The actual channel usage of PU is simulated by queuing system. The probability of PU accessing channel is approximated by Poisson distribution. And the time of a channel being occupied is approximated by geometric distribution. Suppose PU channels are independent to each other. Thus, the parameters of channel state distribution of each channel are also different.


This scenario is illustrated in Fig. 7. Suppose there are $2N$ SUs and in every $T$ time slots, SUs request to access PU channels for communications. Suppose that all of SUs can communicate with each other. Every time a SU will hold $K$ time slots if it succeeds accessing the PU channel. When a SU requests to access a PU channel, the proposed framework makes intelligent decisions based on the results of spectrum prediction and spectrum recommendation to allocate a better PU channel to the SU. However, once PUs need to use the channel again, SUs must release the channel right now. Besides, in this paper it is assumed that when multiple SUs request to access PU channels at the same time, a central node will rank SUs' priority. 
\begin{figure}[!t]
\centering
\includegraphics[width=3.5in]{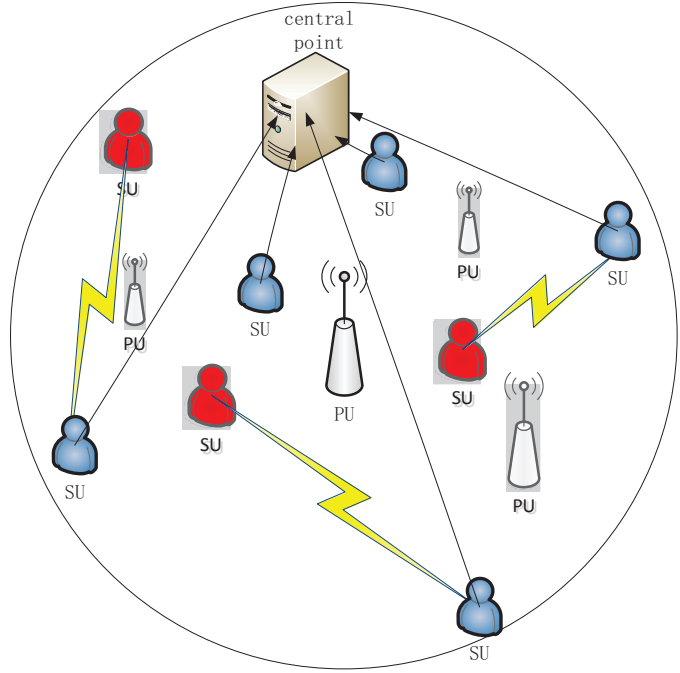}
\caption{Scenario of spectrum decision (I)}
\label{fig:11}
\end{figure}

\textbf{Scenario Two}: SUs' communication distance and locations are considered.

On the basis of spectrum decision system model described in Fig. 8, SU's communication distance is considered. In practice, due to the constraint of hardware equipment and radiation power, node communication distance is limited to a certain range. Suppose each SU has the same communication distance. So, a SU can only communicate with the ones that locate in its communication range. However, when all other SUs that locate within its communication range are under communication, it can not communicate with others and the SU should abandon the request for channel access.

With the consideration of SU's location, the spectrum recommendation algorithm should be modified accordingly. Different locations mean that the similarity between SUs is also different. Therefore, SUs at different locations contribute different weights in the process of channel scoring. Specifically, the shorter distance between SUs, the higher similarity and greater weight they have.
\begin{figure}[!t]
\centering
\includegraphics[width=3.5in]{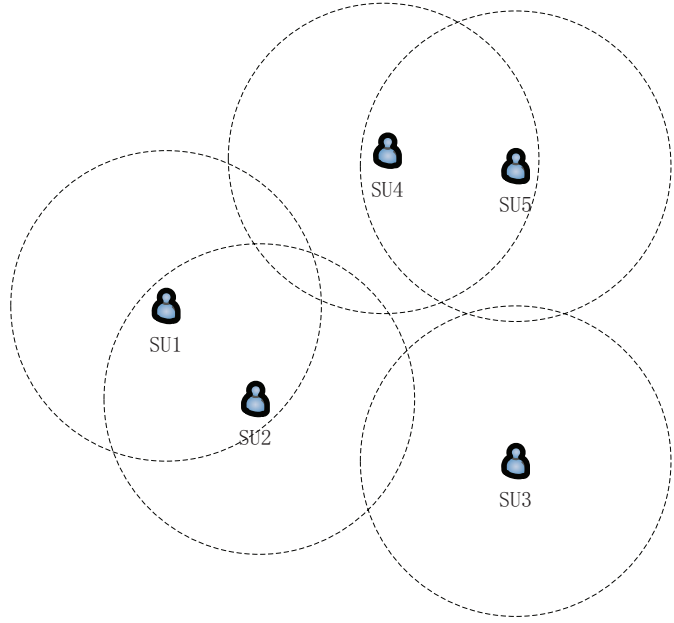}
\caption{Scenario of spectrum decision {II}}
\label{fig:12}
\end{figure}

\subsection{Multiple-Agent system}

From the scenarios introduced in previous section, it can be seen that the whole system involves more than one SU. Therefore, multiple-agent problems are considered in this paper.

Multiple-agent system refers to a system composed of multiple learners, and each agent cooperates with others to complete a task that a single agent can not do. In multiple-agent system, agents may be heterogeneous. The whole system can be affected by the actions that each agent makes. Therefore, the cognitive radio spectrum decision system mentioned above can be regarded as a multiple-agent intelligent learning system. Multiple-agent system is generally divided into types, i.e., single-agent independent learning and multiple-agent cooperative learning. In this paper, we use single-agent independent learning system.

Single-agent independent learning system refers to that each learner in the system is independent in learning process and is not affected from other agents. Each agent can only acquire knowledge and decision information by communication, interaction, feedback, or imitation.

\section{Spectrum Decision Modeling Based on Prediction and Recommendation}

In this section we will introduce the whole process of spectrum decision based on spectrum prediction and spectrum recommendation using Q-learning and Markov decision process (MDP). Q-learning and MDP are two reinforcement learning methods. Their models are almost the same except for some parameters and solutions. So we mainly describe the spectrum decision model based on Q-learning algorithm. The difference between Q-learning and MDP is also described.

\subsection{Spectrum Decision Modeling Based on Q-Learning}

Dynamic spectrum decision method based on Q-learning is illustrated in Fig. 9. And the following is a specific description of the process.

\begin{figure}[!t]
\centering
\includegraphics[width=3.5in]{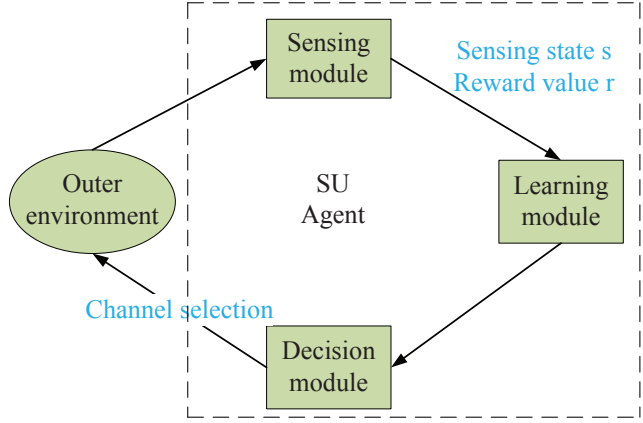}
\caption{Diagram of dynamic spectrum decision using Q-learning}
\label{fig:10}
\end{figure}

\textbf{Environment state $S$}. Suppose there are $M$ PU channels. Each PU channel state is denoted as $d_{i,t}$, which represents PU channel in $i$ state at the time $t$, where $i \in 1,2,...N, d_{i,t} \in \{0,1\}$. 0 means idle state and 1 means busy state. Therefore, $M$ PU channel state values compose a $M$-bit binary integer.
Here $s_{t} \in \{0,1,2,...,2^N-1\}$ and the formula is as follows. All the possible values of $s_{t}$ constitute a set of environment state $S, S = \{0,1,2,...,2^N-1\}$.

\begin{equation}
 S_t=\sum\limits_{i=1}^M d_{i,t}\cdot 2^{i-1}
\end{equation}

\textbf{Action set $A$}. Let $a_{i},i \in \{1,2,...,M\}$ denote the channels that SUs currently can choose to access. Two main factors constrain the choice of system actions. One is whether PU will use channel or not, that is, SUs can only access channel when it is detected to be idle to avoid interfering with PU communications. The other factor is other SUs' actions. That means SUs should choose the channel as many as possible that other SUs have not chosen, which further deducts the probability of collision between users.

\textbf{Reward function $r$}. Reward value is also known as immediate return. When applied to spectrum decision, reward function is designed in this paper to deduct the probability of collision with users and improve spectrum efficiency and system throughput. Since both spectrum prediction and spectrum recommendation can deduct the probability of interference to PUs from the aspects of PUs and SUs, respectively. Therefore when designing the reward function, we mainly consider whether SUs can complete communication without collision with PUs. Meanwhile, the results of spectrum prediction and spectrum recommendation also affect the value of reward function. Specifically, the reward function is set bellow. Let $A$ denote prediction result, $A \in \{0,1\}$, and $B$ denote whether PU channel is in the list of recommendation channel. Specifically, $B=1$ means ``in'' and $B=0$ means not.

When SU completes communication in the selected PU channel without collision with PU, reward function $r$ is set bellow.
\begin{equation}
r = \left\{
     \begin{array}{lcl}
     {300  \qquad A = 0, \ B = 1}\\
     {200  \qquad A = 0, \ B = 0}\\
     {200  \qquad A = 1, \ B = 1}\\
     {100  \qquad A = 1, \ B = 0}
     \end{array}
     \right.
\end{equation}

When SU is in collision with PU before completing communication in the selected PU channel, reward function $r$ is set bellow.

\begin{equation}
r = \left\{
     \begin{array}{lcl}
     {-300  \qquad A = 0, \ B = 1}\\
     {-200  \qquad A = 0, \ B = 0}\\
     {-200  \qquad A = 1, \ B = 1}\\
     {-100  \qquad A = 1, \ B = 0}
     \end{array}
     \right.
\end{equation}

\subsection{The Steps of Spectrum Decision by Q-Learning Algorithm}

\begin{figure}[!t]
\centering
\includegraphics[width=3.5in]{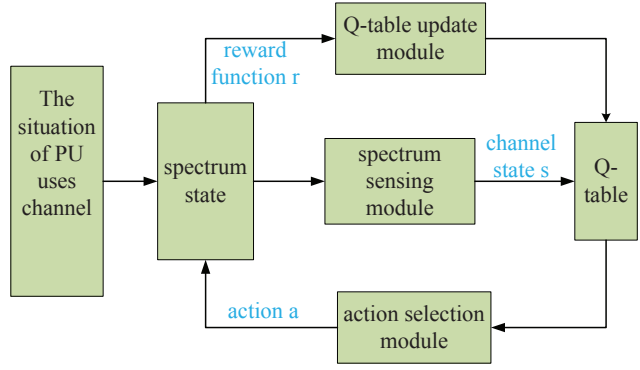}
\caption{ Structure of spectrum decision method based on Q-learning algorithm}
\label{fig:11}
\end{figure}

Fig. 10 shows the structure of the spectrum decision method based on Q-learning algorithm. The proposed algorithm is summarized bellow.
\begin{algorithm}[htb]
\caption{Proposed spectrum decision algorithm based on Q-learning.}
\begin{algorithmic}[1]
\STATE Algorithm initialization. Initialize Q-table, discount factor $\gamma(0\leq \gamma \leq 1)$, and learning rate $\alpha(0\leq \alpha \leq 1)$.
\STATE Constitute state space S. Before requesting to use PU channel, SU detects the PU channel via spectrum sensing. Each PU channel state constitutes current environment state space. Calculate $S_{t}$ as shown in Eq. 17.
\STATE Calculate Q-table. According to current state space $S$, calculate each SU idle channel Q-value.
\STATE Action selection (spectrum decision). Use Eq. 20 to select action $a^{t}$ when $Q$-value is maximized under current state.

\begin{equation}
 a^{t} = \arg \max{(Q^t(s^t,a^t))}
\end{equation}

\STATE Get reward. After SU accesses the selected channel, the reward $r$ is calculated using Eq. 18 and Eq. 19.
\STATE Update $Q$-table. According to the selected action $a$ and reward $r$, $Q$-table is updated.

\begin{equation}
\begin{aligned}
  Q(s_t,a_t) &= Q(S_t,a_t)+ \alpha[r_{t+1} \\
  &+\gamma\max Q(s_{t+1},a_{a+1})-Q(s_t,a_t)]
\end{aligned}
\end{equation}

\end{algorithmic}
\end{algorithm}

When considering SU's communication distance and location, 
we firstly randomly generate each SU's location within a certain area. In the process of generating the channel recommendation list for target SUs, since each SU's location is different, the weights of a $channel_j$ score are also different. Generally, the closer to the target SU, the greater the weights are. For target user $k$, the following equation presents each SU's average scores in recent $L$ on PU channel $j$. And this will be regarded as target user estimation scores on this PU channel.

\begin{equation}
  finalscore_{kj}=\frac{1}{Total_t} \sum \limits_{i=1}^{Total} R_{ij}e^{-d_{ik}}
\end{equation}
where $d_{ik}$ denotes the distance between SU $i$ and target user $k$. $e^{-d_{ik}}$ represents the impact of location on similarity. The farther the distance, the smaller the similarity. 
$R_{ij}$ denotes SU $i$ scores on channel $j$. ${Total}_t$ represents that for current time $t$, the total numbers of SU's accessing channel $j$ during recent $L$ time slots. $finalscore_{k,max}$ is the largest channel scores for current user $k$.

\subsection{The Whole Process of Spectrum Decision}

When a SU wants to communicate, it asks for a request to use PU channel. According to the spectrum sensing process, an idle channel list can be acquired. After determining the current environment state, agent selects the channel that makes the Q-value largest based on Q-table. Spectrum prediction uses history data of spectrum sensing to predict the future channel state. And spectrum recommendation makes recommendation for current SU via other SUs' experiences of accessing channels. Based on current SU's experience and the results of spectrum prediction and spectrum recommendation, a reward value is chosen. Meanwhile, after using the channel acquired by the spectrum decision method, SU will score the channel based on the its usage experience. As a result, the spectrum recommendation will be updated all the time. Therefore, with the agent self-learning constantly, the system will intelligently make spectrum decisions.

\subsection{Spectrum Decision Modeling Based on MDP Method}

Since MDP is similar to Q-leaning, in this section, we only introduce the difference of MDP based spectrum decision method from Q-leaning based method.

1. The state transition probability

Suppose the environment state is $s$ at time $t$, and the action is $a \in A$. Then, the probability of system transfers state $s^{'}$ in next decision time $t+1$ is $p(s^{'}|s,a)$, which is called system state transition probability in MDP. The transition probability is subjected to the following equation.
\begin{equation}
  \sum \limits_{s^{'}\in S} p(s^{'}|s,a)\le 1
\end{equation}
where $\sum \limits_{s^{'}\in S} p(s^{'}|s,a)=1$, if and only if $s,s^{'}\in S, a\in A$.

However, during experiment the state transition probability $p(s^{'}|s,a)$ is unknown. To solve this problem, we first generate PU channel states, then use the statistical method to calculate the transition probability between each state.

2. The value function

Different from Q-learning, MDP has state transition probability. As a result, the value function of MDP is also different from that of Q-learning.
\begin{equation}
  V^{\pi}(s)=R(s)+\gamma \sum \limits_{s^{'} \in S} p(s^{'}|s,\pi(s))V^{\pi}(s^{'})
\end{equation}
where $V^{\pi}(s)$ is the value function. It can be seen that the agent updates the value function after every decision.

3. The solution methods

The common solution methods for MDP model include value iteration and policy iteration. In our experiment, the value iteration method is chosen to solve MDP.

\section{Experimental Results and Discussion}
\subsection{Experimental Result of ELM and BP Based Spectrum Prediction}

The performance of spectrum prediction mainly reflects in the accuracy and speed of the algorithm. Thus, the following experiments mainly focus on those two aspects.

First of all, generate the channel state of primary user. In this experiment, the average arrival interval for primary user is $t_{inter}=10$ slots, and the time that a channel is occupied by primary user is $t_{serv}=10$ slots.  There are totally 10000 slots. The first 5000 slots are used for training, whereas the remaining data are used as test data.
The parameter settings for ELM and BP neural network are shown in Table 1 and Table 2.
\begin{table*}[!htbp]
\caption{Parameters of the proposed spectrum prediction based on ELM} \centering
\begin{tabular}{c|c}
\toprule
Attributes  &  Value \\
\midrule
The number of nodes in input layer $n$ & 10\\
\midrule
The number of nodes in hidden layer $L$ & 30\\
\midrule
The number of nodes in output layer $m$ & 1\\
\bottomrule
\end{tabular}
\end{table*}
\begin{table*}[!htbp]
\caption{Parameters of the proposed spectrum prediction based on BP} \centering
\begin{tabular}{c|c}
\toprule
Attributes  &  Value \\
\midrule
The number of nodes in input layer $n$ & 10\\
\midrule
The number of nodes in hidden layer $L$ & 50\\
\midrule
The number of nodes in output layer $m$ & 1\\
\midrule
The maximum number of iterations: epoch & 200\\
\midrule
Learning rate: lr & 0.2\\
\midrule
Accuracy: goal & 0.0001\\
\bottomrule
\end{tabular}
\end{table*}

In the training set for BP neural network, the maximum number of iterations is less than 200, and we set the learning rate $lr=0.2$. The initial value and threshold for network are generated by Matlab randomly. According to the BP model trained by above mentioned method, the result of spectrum prediction can be obtained.
\begin{figure*}[!t]
\centering
\subfigure[Simulation results of the spectrum prediction method based on ELM neural network]{
\label{fig:Simulation results of the spectrum prediction method based on ELM neural network}
\includegraphics[width=3in,scale=0.5]{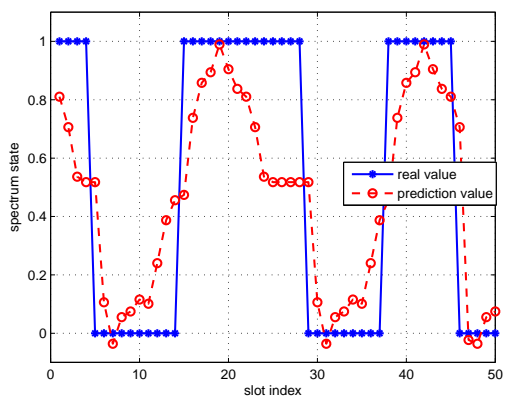}}
\subfigure[Simulation results of the spectrum prediction method based on BP neural network]{
\label{fig:Simulation results of the spectrum prediction method based on BP neural network}
\includegraphics[width=3in,scale=0.5]{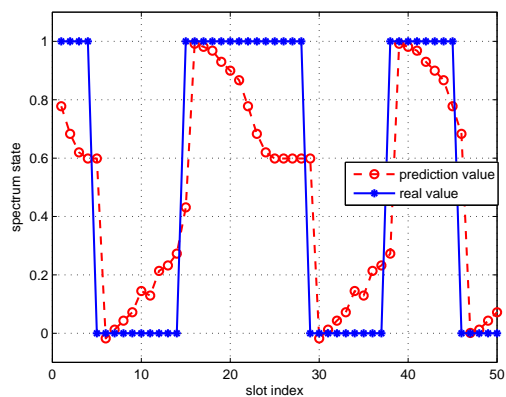}}
\subfigure[Simulation results of spectrum prediction method based on ELM with threshold]{
\label{fig:spectrum prediction method based on ELM with threshold}
\includegraphics[width=3in,scale=0.5]{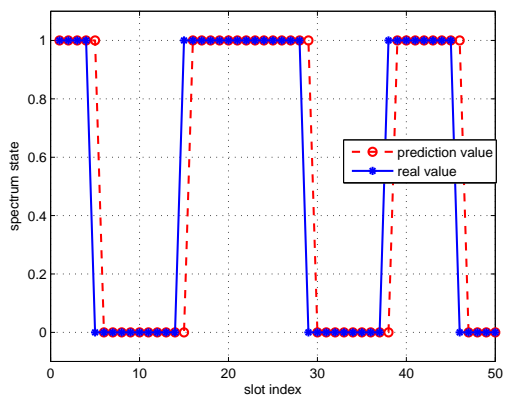}}
\subfigure[Performance comparison of the spectrum prediction methods based on ELM and BP]{
\label{fig:Performance comparison of the spectrum prediction methods based on ELM and BP}
\includegraphics[width=3in,scale=0.5]{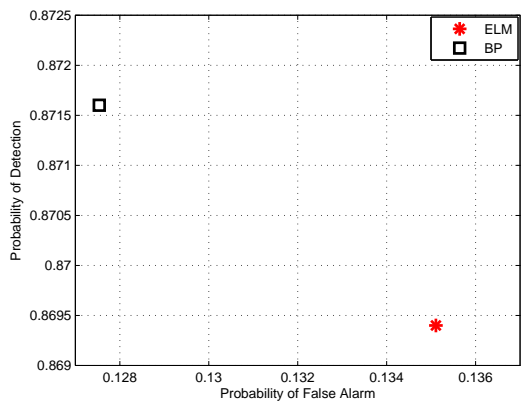}}

\caption{The results of spectrum prediction}

\end{figure*}

Fig.~\ref{fig:Simulation results of the spectrum prediction method based on ELM neural network} and Fig.~\ref{fig:Simulation results of the spectrum prediction method based on BP neural network} show the result of spectrum prediction based on ELM and BP neural network. 
The value of $s_{t+1}^{\alpha}$ can be either 0 (idle channel) or 1 (busy channel). 
Since channel state is generated by mathematical distribution model randomly, and neural network fits the nonlinear law in the training process, so the output of prediction result $s_{t+1}$ is not exactly ``0" or ``1". Thus, a decision threshold $\lambda$ is set to 0.5. 
\begin{equation}
s_{t+1}^{p}\!=\!\begin{cases}
1 & if  s_{t+1}\geq \lambda\\
0 & if  s_{t+1}\leq \lambda\\
\end{cases}
\end{equation}
where $s_{t+1}$ denotes the prediction result of next slot. $\lambda$ denotes the decision threshold. Fig.~\ref{fig:spectrum prediction method based on ELM with threshold} shows the simulation results of spectrum prediction method based on ELM with threshold. From Fig.~\ref{fig:spectrum prediction method based on ELM with threshold} we can also learn that prediction errors mainly occur at the alternation of channel state.

In spectrum prediction, there are two normal parameters that can be used to measure the performance, i.e., probability of detection $P_{D}$ and probability of false alarm $P_{FA}$.
\begin{equation}P_{D}=P(s=1|s^{\alpha}=1)=\frac{\sum_{i=1}^{t_{max}}(s_{i}=1|s_{i}^{\alpha}=1)}
{\sum_{i=1}^{t_{max}}(s_{i}^{\alpha}=1)}\end{equation}
\begin{equation}P_{FA}=1-P(s=0|s^{\alpha}=0)=1-\frac{\sum_{i=1}^{t_{max}}(s_{i}=0|s_{i}^{\alpha}=0)}
{\sum_{i=1}^{t_{max}}(s_{t}^{\alpha}=0)}\end{equation}
\begin{figure}[!t]
	\centering
	\includegraphics[width=3.5in]{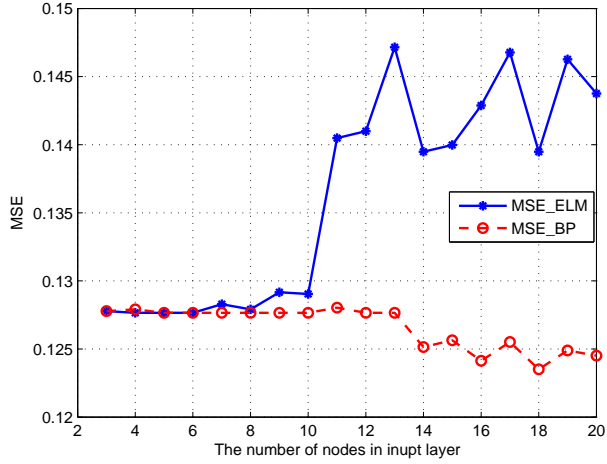}
    \caption{The MSE comparison of the spectrum prediction methods based on ELM and BP with the increase of input layer nodes}
	\label{fig:mse comparison of the spectrum prediction methods}
\end{figure}

The performance comparison of the spectrum prediction methods based on ELM and BP is shown in Fig.～\ref{fig:Performance comparison of the spectrum prediction methods based on ELM and BP} and Fig.~\ref{fig:mse comparison of the spectrum prediction methods}. 
We can learn that BP neural network based spectrum prediction algorithm is better than that of ELM. Fig.~\ref{fig:mse comparison of the spectrum prediction methods} shows with the increase of input layer nodes, the MSEs of the ELM and BP based spectrum prediction. The prediction accuracy of BP neural network is better than that of ELM, whereas the training speed of ELM spectrum prediction algorithm is faster than BP spectrum prediction algorithm. The training time for BP and ELM are 4.4631 seconds and 0.0486 seconds, respectively. According to Eq. 28 and Eq. 29 the training speed increases by $98.92\%$ and training time decreases by 92 times.
\begin{equation}I_{speed}=\frac{t_{BP}-t_{ELM}}{t_{BP}}\times 100\%\end{equation}
\begin{equation}D_{time}=\frac{t_{ELM}}{t_{BP}}\end{equation}
where $I_{sppeed}$ denotes the percentage of increased training speed. $D_{time}$ denotes the times of decreased training time.

\subsection{Experimental Result of Q-learning Based Cooperative Spectrum Prediction}
\subsubsection{M-out-of-N}

In order to evaluate the performance of Q-learning based cooperative spectrum prediction algorithm, the real-world Wi-Fi signals which was measured in \cite{Chen2010tvt} is used as test data. In this section, M-out-of-N cooperative spectrum prediction method is used as contract. Eq. 30 shows the mathematical expression of M-out-of-N cooperative spectrum prediction.
\begin{equation}
Channel_{state}\!=\!\begin{cases}
1 & if  \sum_{i=1}^{N}s_{i,t}\geq M\\
0 & if  \sum_{i=1}^{N}s_{i,t}< M\\
\end{cases}
\end{equation}

If $M=1$, the algorithm of M-out-of-N is equivalent to the ``OR" rule. And if $M=N$, it is equivalent to the ``AND" rule. In this experiment, the number of secondary user is 3, thus $N=3$. $Channel_{state}$ denotes final result of cooperative spectrum prediction.

The following cooperative spectrum prediction method is proposed in \cite{Chen2010tvt}.
\begin{equation}
Channel_{state}\!=\!\begin{cases}
0, & if  \sum_{i=1}^{N}\frac{P_{0i}-P_{1i}}{P_{0i}+P_{1i}}\geq 0\\
0 & if  \sum_{i=1}^{N}\frac{P_{0i}-P_{1i}}{P_{0i}+P_{1i}}< 0\\
\end{cases}
\end{equation}
where $P_{0i}$ denotes the probability of secondary user $SU_{i}$ predicting the channel state to be ``Idle". And $P_{1i}$ denotes the probability of secondary user $SU_{i}$ predicting the channel state to be ``busy".

\subsubsection{Hidden Markov Model}

A hidden Markov model (HMM) is defined by a tuple $\lambda=\{\pi,A,B\}$, $\pi$ is the initial state probability vector,
\begin{equation}\pi=(\pi_{1},\cdots,\pi_{N})\end{equation}
\begin{equation}\pi_{i}=Pr(q_{1}=\theta_{i})\;
                i=1,\cdots,N
\end{equation}
where $Pr(\bullet)$ denotes probability, N is the number of states of Markov chain, $\{\theta_{1},\cdots,\theta_{N}\}$ are the $N$ states, $q_{t}$ represent the state at time $t$, $A$ is state transition matrix.
\begin{equation}A=(\alpha_{ij})_{N\times N},\alpha_{ij}=P(q_{t+1}=\theta_{j}|q_{t}=\theta_{i}),i,j=1,\cdots,N_{i}\end{equation}
And $B$ is emission probability matrix.
\begin{equation}B=(b_{ij})_{N\times M}\end{equation} \begin{small}\begin{equation}b_{jk}=P(o_{t}=v_{k}|q_{t}=\theta_{i})=b_{j}(o_{t}),i,j=1,\cdots,N,k=1 \cdots M_{i}\end{equation}\end{small}
where $M$ is the number of possible observation values in the observation space $\{v_{1},\cdots,v_{M}\}$, $o_{t}$ represents the observation value at time t, $o_{t}\in \{v_{1},\cdots,v_{M}\}$.
According to a statistic method proposed in \cite{Chen2010tvt}, the state transition matrix $A$ and the emission probability matrix $B$ can be obtained. When the model of HMM based spectrum prediction is built, we can use the following method to do spectrum prediction.
\begin{equation}\delta(i)=\pi_{i}b_{i}(o_{1}),i=1,\cdots,N_{i}\end{equation}
\begin{equation}\delta_{i}(j)=\max \limits_{1\leq i \leq N_{i}}[\delta_{t-1}(i)\alpha_{ij}]b_{j}(o_{t}),j=1,\cdots,N_{i},t=2 \cdots T\end{equation}
\begin{equation}P^{*}=\max \limits_{1 \leq i \leq N_{i}}[\delta_{T}(i)]\end{equation}
\begin{equation}q_{T}^{*}=\arg\max \limits_{1\leq i \leq N_{i}}[\delta_{T}(i)]\end{equation}
$P^{*}$ is the calculated likelihood probability and $q_{T}^{*}$ is the estimated state at time $T$.

\begin{figure*}[!t]
\centering
\subfigure[Performance of cooperative spectrum prediction using measured data]{
\label{fig:Performance of cooperative spectrum prediction using measured data}
\includegraphics[width=3in]{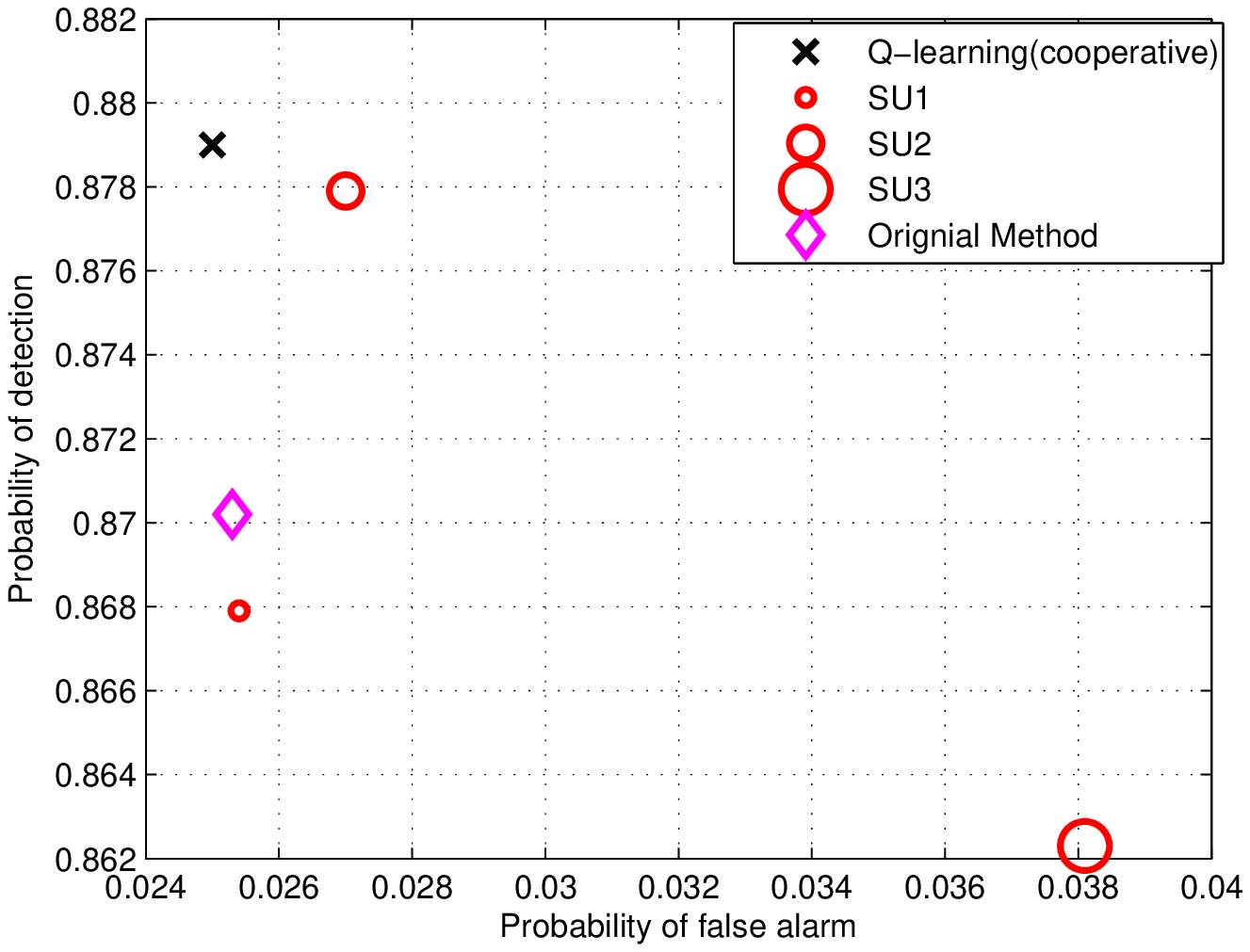}}
\subfigure[The comparison of the performance for cooperative spectrum prediction methods]{
\label{fig:Performance contrast of cooperative spectrum prediction methods using measured data}
\includegraphics[width=3in]{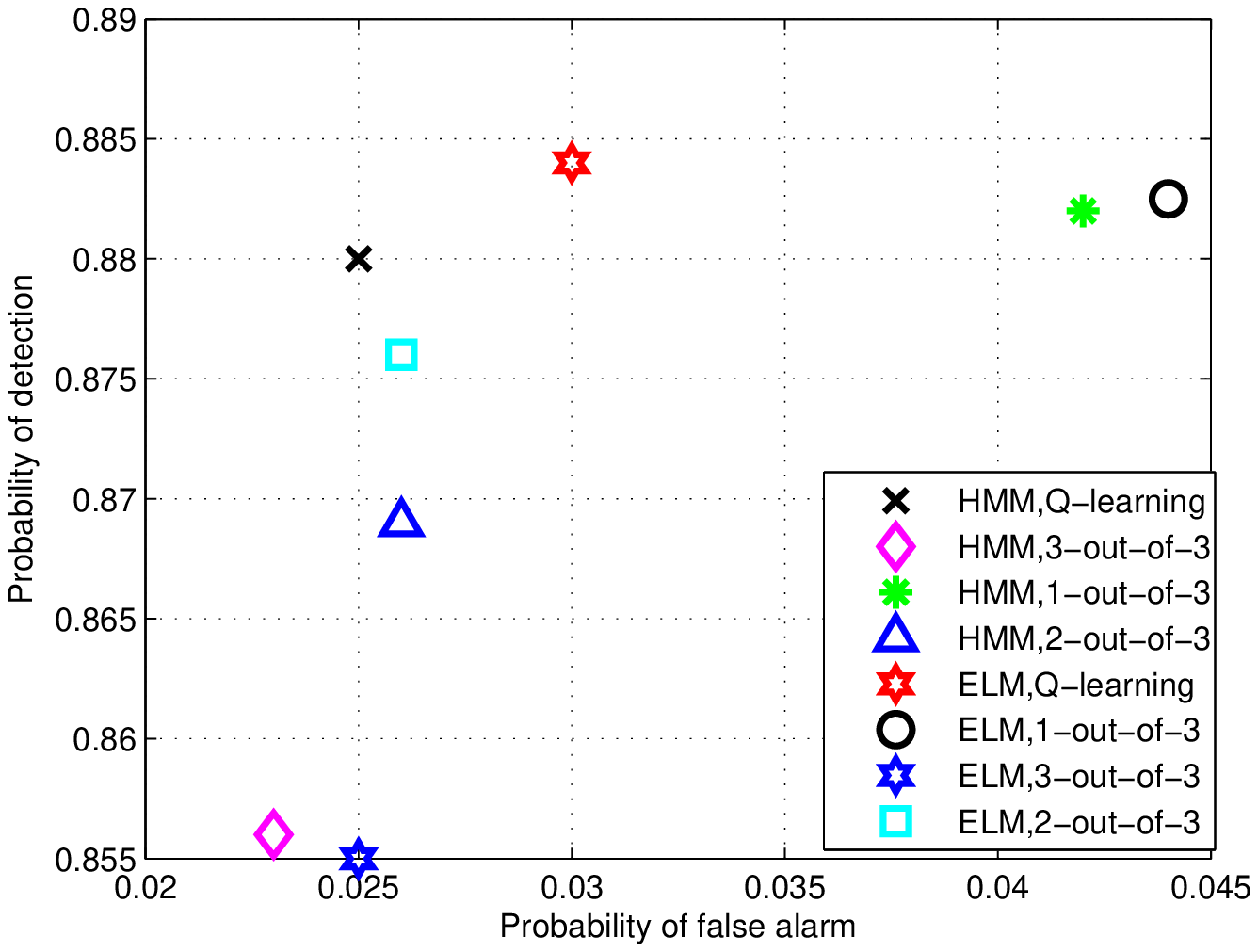}}
\caption{The result of cooperative spectrum prediction}
\end{figure*}

Fig.~\ref{fig:Performance of cooperative spectrum prediction using measured data} shows the performance of cooperative spectrum prediction using measured data. And we can learn that the accuracy of Q-learning based cooperative spectrum prediction is higher than the prediction of independent secondary users. Compared with soft decision cooperative method, Q-learning based cooperative spectrum prediction is more accurate.

From Fig.~\ref{fig:Performance contrast of cooperative spectrum prediction methods using measured data} we know that if $M=1$ or $M=3$ the performance is not very good. And $M=2$ is better. 
Taking into account of $P_{FA}$ and $P_{D}$, we can also learn that Q-learning based algorithm proposed in this paper outperforms M-out-of-N.

\subsection{Experimental Result of Collaborative Filtering Based Spectrum Recommendation}

The channel state of each primary user are independent from each other, where the average channel occupance time $t_{serv}$ of the primary user in the first 4 channels is $\lambda_{1}$ slots, which is generated randomly from 1 to 10. The average arrival interval $t_{inter}$ of primary user is $\lambda_{2}$ slots, which is generated randomly from 10 to 20. And the last channel is set to be ``idle". The score for secondary user is from 0 to $K$ and it is determined by the number of slots of successful transmissions.
The threshold is set to $Th=finalscore_{max}/2$.

When secondary user accesses channel, collision may occur if other user comes in while the secondary user is in transmission. The collision rate $P_{collision}$ and the average number of successful communication per $T$ time slots $D_{e}$ is defined as follows.
\begin{equation}
    P_{collision}=\frac{N_{collision}}{N_{total}}
\end{equation}
\begin{equation}
D_{e}=\frac{D_{success}}{N_{total}}
\end{equation}
where $N_{collision}$ is the number of collisions. $N_{total}$ is the total number of access channel. $D_{success}$ is the number of successful transmission.
\begin{figure}[!t]
	\centering
	\includegraphics[width=3.5in]{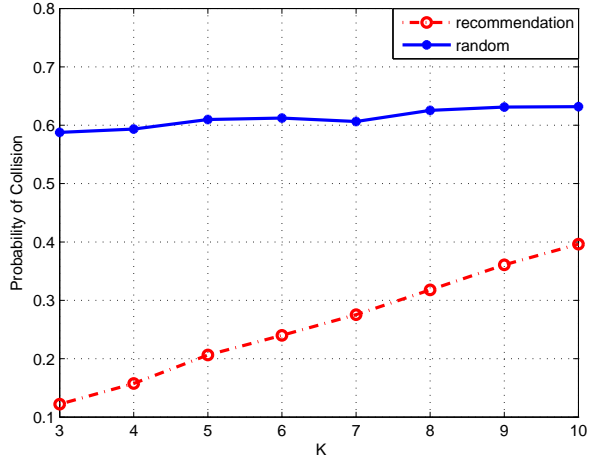}
    \caption{Simulation results of the probability of collision of spectrum recommendation}
	\label{fig:Simulation results of the probability of collision of spectrum recommendation method}
\end{figure}
\begin{figure}[!t]
	\centering
	\includegraphics[width=3.5in]{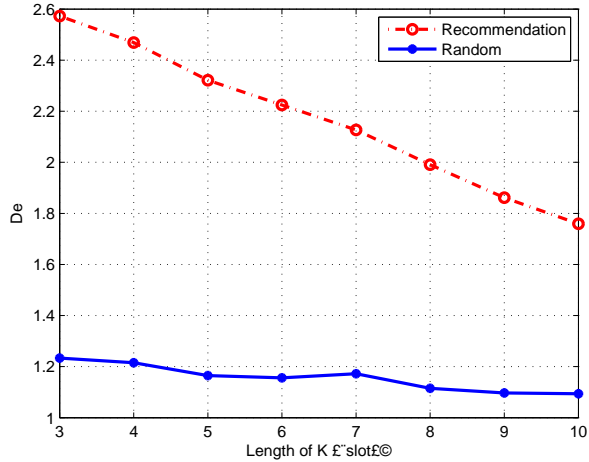}
    \caption{Simulation results of the number of successful transmissions during $T$}
	\label{fig:Simulation results of the number of successful communication during T}
\end{figure}

Simulation results of the probability of collision of spectrum recommendation is shown in Fig.~\ref{fig:Simulation results of the probability of collision of spectrum recommendation method}. And from Fig.~\ref{fig:Simulation results of the probability of collision of spectrum recommendation method} we can know that the collision rate is decreased while using collaborative filtering based spectrum prediction method. Fig.~\ref{fig:Simulation results of the number of successful communication during T} shows the simulation results of the number of successful transmissions during $T$. 

In conclusion, cooperative filtering based spectrum prediction method can be used to select channels. It can decrease the collision rate and improve the spectrum utilization.

\subsection{Experiment Results of Spectrum Decision Based on Prediction and Recommendation}
\subsubsection{Scenario One}

\begin{figure}[!t]
\centering
\includegraphics[width=3in]{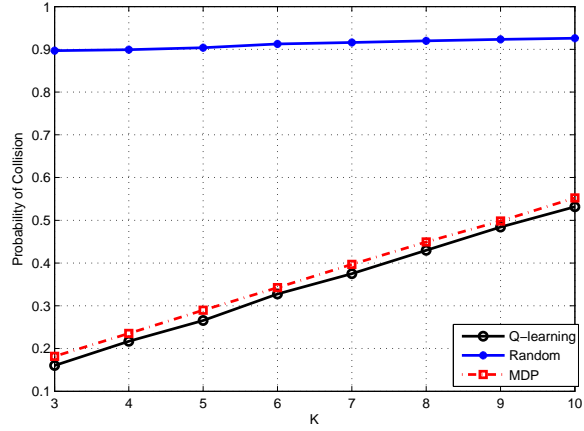}
\caption{Collision probability of the proposed spectrum decision method (in scenario one)}
\label{fig:24}
\end{figure}

\begin{figure}[!t]
\centering
\includegraphics[width=3in]{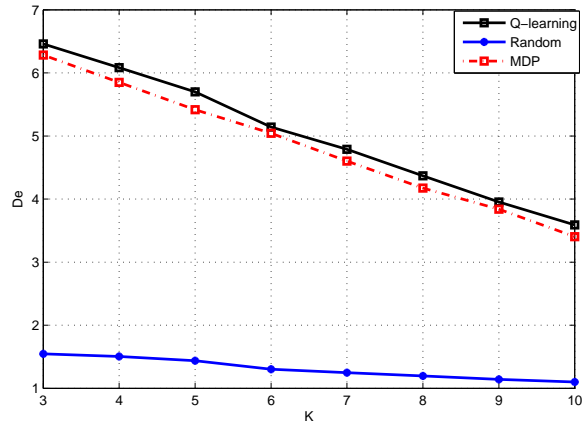}
\caption{Number of successful transmissions during $T$ (in scenario one)}
\label{fig:25}
\end{figure}

In this section, the proposed spectrum decision method is compared with random access. By random access, we assume that there is a central node that randomly allocates a current idle channel to SUs.

In simulation, the number of SU is $N=30$ and the number of PU channel is $M=10$. 
The first 9 channels are averagely occupied by PUs for $\lambda_1$ time slots. The $\lambda_1$ is randomly generated in the range of 1 to 10. The average interval of PU is $\lambda_2$, which is randomly generated in the range of 10 to 20. The last PU channel is set to be idle. In this simulation, there are totally 1000 time slots.

In spectrum recommendation, after using PU channels for transmission, SUs share their experiences with each other by scoring the channels. The scores are decided by the number of slots that SUs succeed to transmit. In the generation of the recommendation list, let latest $L=10$ scores be the reference and set the threshold $Th=finalscore_{k max}/2$.

For spectrum prediction, the spectrum prediction method based on ELM is employed. 

For Q-leaning, set $\gamma=0.5,\alpha=0.5$. In every time slot, there are SUs' requests to access channels. When accessing channels, SUs transmit for $K$ time slots. Here, $K$ is a variant. In this experiment, $K$ is set to be $K=3,K=4,...,K=10$ in turns. Let $T=K$.


The figures show that compared with random access, the proposed spectrum decision method based on both spectrum prediction and spectrum recommendation greatly reduces the probability of collision between users and increases the number of successful transmissions during $T$, which improves the utilization of the spectrum. In our spectrum decision method, with longer transmission time, the collision probability between users increases a little. This is identical to the practice. However, when SU's transmission time is determined, our spectrum decision method works well. Due to the excellent self-learning ability, Q-learning and MDP based spectrum decision methods can continuously learn and adjust the spectrum selection policy under various channel state, based on the information of spectrum prediction and spectrum recommendation. As a result, the channel with less probability of collision and higher rating scores is allocated to SU. Besides, in the case of more than one SUs selecting the same channel, under the control of a central node, the probability of collision can be reduced.

It also can be seen that compared with MDP based spectrum decision method, the Q-learning based spectrum decision method performs slightly better. This is because that in our experiment, the state transition probability is calculated using a statistical method. However, due to the limit amount of data, the calculated state transition probability may not exactly reflect the actual state transition situation, which affects the convergence of the MDP based spectrum decision algorithm.

\subsubsection{Scenario Two}

\begin{figure}[!t]
\centering
\includegraphics[width=3in]{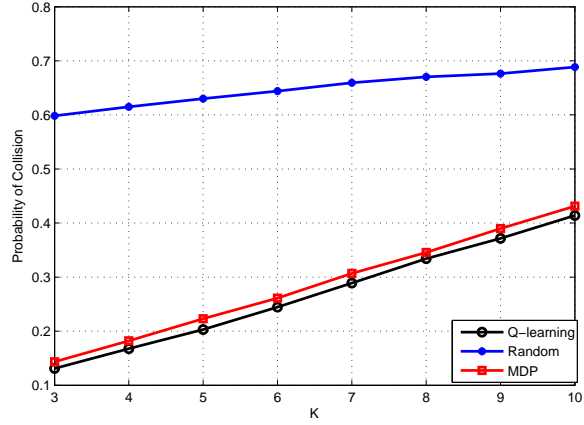}
\caption{Collision probability of the proposed spectrum decision method (in scenario two)}
\label{fig:26}
\end{figure}

\begin{figure}[!t]
\centering
\includegraphics[width=3in]{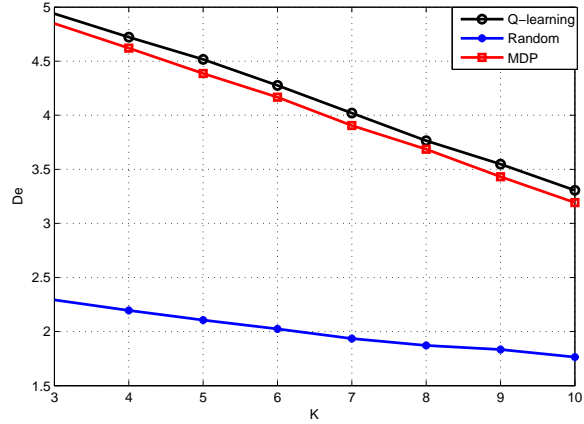}
\caption{Number of successful transmissions during $T$ (in scenario two)}
\label{fig:27}
\end{figure}

In this section, SUs' communication distance and locations are considered. Except the setting of SUs communication distance and locations, the parameters are identical to those in scenario one.

In a $40\times 40$ simulated space, the position coordinates of $N=30$ SUs are generated randomly. Each SU's communication range is a circle of radius $r_i=5$. Each SU can only communicate with others located in its communication range. In the process of this experiment, some SUs are randomly selected to request communications with other SUs within their communication range. If all other SUs located within its communication range are not available, the SU should abandon request to access the channel. In each time slot, the number of SUs requesting to access channel is uncertain. According to the results spectrum decision, SUs select better idle channels for transmission and judge whether a collision is occurred based on the actual channel states.

The simulation results are shown in Fig. 18 and Fig. 19. It can be seen that, compared with the results of scenario one, the overall probability of collision between users and the number of successful transmission during $T$ is reduced a little with the consideration of communication distances and locations. Apparently, this is because the number of SUs that can communicate during $T$ is reduced. But the whole experiment results are similar to those in scenario one. This experiment also demonstrates that the proposed framework of spectrum decision based on spectrum prediction and spectrum recommendation can reduce the probability of collision between users and improve the utilization of the scarce spectrum resources.

\section{Conclusion}

This paper aims to maximize the overall spectrum utilization and minimize collisions with primary users for cognitive radio. Spectrum decision plays an important role towards this goal. In this paper, a framework for spectrum decision based on spectrum prediction and spectrum recommendation has been proposed. Moreover, for spectrum prediction, a prediction method based on extreme learning machine (ELM) for single-user spectrum prediction and a method based on Q-learning for multiple-user spectrum prediction are proposed. And two methods based on Q-learning and Markov decision process (MDP) are also proposed to enhance the overall performance of spectrum decision. Experimental results show that the proposed spectrum decision framework is feasible and effective.

\section*{Acknowledgment}

This work is supported by the Fundamental Research Funds for the Central Universities (N140404015).
Mr. Sai Xie and Mr. Yongshuai Shao make a major contribution to the English writing of this paper.

\bibliographystyle{ieeetr}
\bibliography{cr}
\end{document}